\documentclass[]{article}
\usepackage{graphics}
\usepackage[varg]{txfonts} 
\usepackage[latin1]{inputenc}
\begin{document}
\title{Optimal control of laser plasma instabilities using Spike Trains of Uneven Duration and Delay (STUD pulses) for ICF and IFE}
\author{Bedros Afeyan\footnote{email: bafeyan@gmail.com}  \\ 
Polymath Research Inc., Pleasanton, CA, 94566, USA  \\
Stefan H\"uller \\ 
Centre de Physique Th\'eorique, CNRS, Ecole Polytechnique, Palaiseau, France}
\maketitle
\abstract{
An  adaptive method of controlling parametric instabilities in laser produced plasmas is proposed. It involves fast temporal modulation of a laser pulse on the fastest instability's amplification time scale, adapting to changing and unknown plasma conditions. These pulses are comprised of on and off sequences having at least one or two orders of magnitude contrast between them. Such laser illumination profiles are called STUD pulses for Spike Trains of Uneven Duration and Delay.  The STUD pulse program includes scrambling the speckle patterns spatially in between the laser spikes. The off times allow damping of driven waves. The scrambling of the hot spots allows tens of damping times to elapse before hot spot locations experience recurring high intensity spikes. Damping in the meantime will have healed the scars of past growth. Another unique feature of STUD pulses on crossing beams is that their temporal  profiles can be interlaced or staggered, and their interactions thus controlled with an on-off switch and a dimmer. }


%
%
\vspace*{-.5cm}
\section{Introduction} \label{intro}
\vspace*{-.2cm}
A method was conceived in April-May 2009 \cite{STUD} to optimally control the
unbridled growth of parametric instabilities in laser-produced plasmas. It was a
response to the 
question: How could one keep parametric
instabilities in the strictly linear regime, or any other well specified regime,
for that matter? This led to the realization that the answer must lie in the manipulation of the laser
pulse and not in the plasma properties which are hard to know and much harder
to change. It was further clear that a spike train of isolated pulses would
be required and not a continuous illumination scenario. The reasons for this
had to do with adaptivity. Adapting to the changing plasma conditions, to
modulated beams influencing the plasma, influencing the instabilities of
other beams, would be a key feature. The idea was to gauge how many
growth cycles had gone by, by monitoring reflectivities on a very fast time
scale, and stop the growth at a fixed number. Then, after a time of relative
calm, the process could be started again.  
At the same time, questions were being raised on the LPI problems that might arise in
the Mourou proposed million fiber laser march to inertial
fusion energy (IFE)  \cite{Mourou}.  The 
superposition of such a large set of beams made it inevitable that if STUD 
pulses were used in that million or more fiber scheme,
the speckle patterns would constantly change from spike to spike and that the
sustained local growth in laser hot spots could be tamed as long as the damping
time was short compared to the recurrence time of hot spots, which would favor
the strong damping limit. These analyses pointed  to the need for on-off pairs to comprise
the fundamental building blocks of future  IFE and inertial confinement fusion (ICF) laser illumination
schemes. Then, it also became clear that with so many lasers overlapping on 
target, staggering their pulses in time, or interleaving them, would lead to 
further control of their mutual interactions. That gaps would be needed in one
pulse to accommodate a set of different beams with staggered pulses, became the
paradigm of the STUD pulse program. The major elements of STUD pulse physics can be highlighted as follows:

(1)  Solve the LPI problem by engineering the laser pulse on the ps time scale \cite{Weiner}. One can have far stronger control over laser pulse shapes than  on plasma conditions which one cannot even measure in detail as they evolve and change due to multiple other interactions occurring simultaneously. STUD pulse designs point to a path of LPI control that does not rely on any prior knowledge of the evolving plasma conditions inside a hohlraum or a direct drive target. It relies on trying on different pulse shapes to back out the fastest growing instability conditions inside the plasma and how to tame the instabilities accordingly. Also, novel measurement schemes become possible using pump-probe systems using STUD pulses of the kinetic state of a plasma.

(2) Turn the laser on and off at a rate which does not allow sustained instability growth at particular positions where laser hot spots may be present. Scramble the hot spots so that gain will become locally halted and spatially globally distributed and not be continuously  reinforced for nanoseconds (see (7) below for a spatial perspective on this point). Also note that this is far more effective in 3D than 2D, to which our numerical work has so far been limited. 

(3) Let the waves damp in between significant recurrence events. This is when two consecutive complex Gaussian random field samples \cite{Adler} happen to have hot spots within a half length and a half width of each other and with a small range of intensity separating them. This is an exponentially small probability event for the hottest hot spots \cite{Azais} and so seldom more than $10\%$ of the hot spots will be subjected to accidental recurrence twice in a row, and far less chance to hit the
same set three or more times in a row.  If the waves damp before high intensity revisits an area where large growth has already occurred, then it is as if (almost) nothing had happened before. So coherent, persistent growth becomes replaced by incoherent sums of growth spurts in isolated areas of space and time. 

(4) Interleave crossing beams in time at the expense of making their peak intensities higher to compensate for the off times, so that the overall energies in the beams are held fixed.  Increasing peak intensities could exceed glass damage thresholds and frequency tripling conversion efficiency issues.  A compromise has to be sought between taming LPI and having a laser to drive the implosions correctly on the hydrodynamical time scale. 

(5) STUD pulses include the ISI scheme \cite{Lehmberg} as a natural reduced limit. ISI itself is far superior to other beam smoothing techniques such as RPP and SSD. The latter are discussed in a companion paper by H\"uller in this volume and compared to STUD pulses and found wanting \cite{ifsa-stefan}. 

(6) We distinguish between various STUD pulse designs. First there is the duty cycle. That is, the "on" time, divided by the total time of  an elemental on+off  pair. Distinguished limits are $20\%,$ $50\%,$ and $80\%.$ In addition, a random jitter is best applied to the spike widths (and heights to keep the height$\,\times\,$width product constant). The purpose is to avoid resonances between harmonics contained in the sharp edges of the spikes on the one hand and slow waves in the plasma such as ion acoustic waves (IAW). By adding a $10\% $ of "on" time jitter to a pulse, this tendency can be kept in check. For $50\%$ duty cycle $+10\%$  jitter we arrive at the notation STUD $\!5010$ pulses. Figure 1 shows samples of STUD pulses in the three duty cycle distinguished limits but with the same fundamental modulation rate. One more independent crucial element in specifying a particular STUD pulse is the number of spikes before the RPP patterns are scrambled. If they are not changed at all and a static RPP pattern is used throughout the pulse, we arrive at  a STUD $5010\!\times\infty$ pulse. The best STUD pulses will be in the opposite limit having the $\times 1$ suffix as opposed to $\times2$ or $\times3, $ or indeed, $\times\infty.$

(7) There is a spatial-length-scales comparison method of arriving at the optimal STUD pulse modulation rate for a given set of plasma and average laser intensity conditions. In the convective instability regime there are three significant length scales that can be ordered to advantage. One is the hot spot length, $L_{HS} \approx 4 f^2 \lambda_0,$ \cite{speckle-stat}. Another is the interaction length of the convective instability (which scales differently in the weak and strong damping limits), $L_{INT},$ (see the next section for explicit details) and the third is the distance traveled by the scattered light wave during the typical "on" time of a STUD pulse, $L_{spike}  \approx c \times t_{spike}.$ The most interesting regime is when $L_{HS} \approx L_{INT}$ and $L_{spike}$ is half as long. This is the so called "cut the interaction length in half" regime. Starting with longer pulses, when this criterion is met, strong improvements in the performance of STUD pulses will kick in. If $L_{INT} \ll L_{HS},$ then scrambling hot spots by changing the speckle patterns is even more effective, since hiding previously amplified tiny portions of hot spots becomes proportionately easier. But then not much reflectivity can be expected in that case.  This is due to much shortened gain lenghts and a push towards the very hottest hot spots dominating the reflectivity. But due to very little energy content of these hot spots (and pump depletion) their reflectivity can never be too high due to exponential statistics. If, instead, $L_{INT} \gg L_{HS},$ then scrambling the hot spot patterns will not be as effective. There are a number of phase transitions of this intricate sort lurking behind this problem signaling different regimes of collective behavior and global self-organization which we have only hinted at here and which we hope to elucidate in great detail in future publications. 

(8) The window on modulation rate to tame high frequency instabilities is not wide. It is not productive to modulate a laser pulse at a rate short compared to the lifetime of the lasing line. That puts a hard limit at 500-700 fs for glass lasers. It is also ill advised to modulate the laser at a rate which changes or strongly interferes with the hydro. That puts a hard limit at 50 ps. So the real window available is from 1-2 on up to 10-20 ps for a baseline modulation period to execute the STUD pulse program to effectively control Raman backscattering on the fast end and Brillouin scattering and crossed beam energy transfer on the slow end.
\vspace*{-.5cm}
\section{Analytic Estimates and Simulation Results}
\vspace*{-.2cm}

STUD pulses may be implemented as a sum of localized super-Gaussian spikes with a super-Gaussian exponent of 8-10. The gaps between the spikes must be preserved so that merging of successive spikes does not occur. Fig. 1 shows three instances of such pulse shapes. Note the random modulations in pulse width of the order of 10\% that have been added with a concomitant change in the amplitudes of the spikes so that their width $\!\times\!$ height product is conserved. This is not necessary but it is a convenient way of constructing STUD pulses from elementary constitutive parts. The contrast between the on and off portions is also an important design parameter for STUD pulses. It can be used to define the intensity at the "off" sections of the pulse. It is recommended that this contrast not be much less than two orders of magnitude so as to allow temporal interlacing of different beams to be effective. Too small a contrast and the gaps will get filled in and instabilities will grow at much higher rates with no damping and healing and democratization via an incoherent sum of growth spurts, Without these features of STUD pulses instabilities tend to self-organize and establish long range order as in the case of RPP or SSD. Note that for convective instabilities, the gain is the same whether one is in the strong or weak damping limits. It is the interaction length over which that gain occurs that increases in the case of strong damping.

For the sake of concreteness, stimulated Brillouin backscattering (SBS) is treated in this section. The physical principles and technical execution of the STUD pulse program are the same for SRS or SBS. The time scales are not quite the same and transitions to the strong coupling regime is more of an issue for SBS than SRS. Indeed an inhomogeneously flowing plasma profile brings with it complications not found in inhomogeneous density profiles, such as in SRS. On the other hand, kinetic nonlinearities tend to affect SRS much more readily than SBS. Furthermore, complications at Mach $-1$ and Mach $0$ surfaces and nonlocal resonances between strongly coupled SBS growth in intense hot spots and weak coupling growth elsewhere at a different flow speed, make for rich physics in static or slowly varying laser profiles such as RPP or SSD, as shown in the companion paper by H\"uller, which STUD $\!5010\!\times\!1$ pulses then help eliminate. How the transitions occur as a function of duty cycle, modulation rate and hot spot scrambling rate are shown here. The run parameters and the description of the numerical model used can be found in \cite{ifsa-stefan}.  A tenth critical plasma at 2 keV having a thousand hot spots with f/8 optics in the Green at an intermediate damping regime, where the damping rate divided by the canonical IAW frequency is 0.025, are the general conditions where the intensity is chosen such that the convective gain at the lowest velocity (Mach -3) at the average intensity is $\approx 5.5.$ The flow profile in terms of Mach numbers extends from -3 to -7 in the physical region of the simulation box. There will be 8 to 10 hot spots axially and hundreds of hot spots in the lateral direction with roughly equal unoccupied regions in between. 

The interaction length for SBS in the convective instability regime in the strong damping limit, in physically accessible units is given by:
$$ L_{INT, SBS, 100\mu m}^{SDL} = 0.2 \, [ | 1 + 1 / M| ] \times L_{v, 100\mu m} \times (\nu_{IAW} / (0.2 \, c_s \, k_0) ) . $$
Here $L_{v}$  is the velocity gradient scalelength at the lowest velocity reference point where the Mach number is M and the gain is largest. The SBS Rosenbluth gain exponent itself is given by:
$$G_{MNR}^{SBS} = 9.2 \,  (n/n_c) \, ( I_{14 W/cm^2} \lambda_{0, \mu m} / T_{e,keV} ) \times (L_{v, 100\mu m}  / | M|).$$

These translate to simulation parameters as given above and elaborated upon further in the companion paper by H\"uller \cite{ifsa-stefan}.

These are not ideal conditions especially since the damping regime is also an intermediate one between strong and weak. So the results are not typical. They overplay the hand of STUD pulses by making the interaction length much smaller than the typical hot spot length thus making scrambling of the hot spots much more effective than  can be expected, but underplay the role of damping as a healer since the rate is four times lower than expected for hydrogen or helium  plasmas. Additionally, the modulation rate for the STUD $\!5010$ cases is too slow by almost a factor of two and the interaction length is not being cut in half either.  Better parameter regimes are being explored and will be reported soon. Note that STUD $\!8010$  very nearly approaches an ISI beam at a particular bandwidth (which in STUD pulse language becomes the speckle pattern resampling or scrambling rate). The point of showing STUD $\!8010$ cases here that have twice the modulation frequency of STUD $\!5010$ cases shown, is to see if a faster STUD $\!8010\!\times\!1$ could catch up with a slower modulated STUD $\!5010\!\times\!1$ pulse's performance. The answer is yes at  $\!\times1$ but not at $\times3$ or $\times\infty.$  STUD $\!8010$ pulses, having very little off time, do not allow much damping and healing to occur before recurrences. This is especially so when the recurrence is forced by waiting longer periods between the scrambling of hot spot patterns such as $\times3$ and $\times\infty.$ Most importantly, ISI or STUD $8010$ pulses do not have room for interlacing of different beams in time and thus cannot be advised as proper solutions when many overlapping laser beams must be deployed to couple energy to a target simultaneously. STUD $\!2010\!\times\!1$ pulses offer much more flexibility there, at the cost of five times higher peak intensity than the average. 

Below, we show reflectivity results from $\!\times\!1, \times3$ and $\times\infty$ cases of STUD $\!5010$ and STUD $8010$ in a case where $L_{INT} \approx L_{HS} / 5$ and where $L_{spike} \approx L_{HS}.$  Samples of these three STUD pulse shapes are shown in Fig. 1.  Figure 2 shows the SBS reflectivity of  STUD $\!5010\!\times\!1$ pulses together with those of STUD $\!5010\times\!3$ and STUD $\!5010\times\!\infty,$ compared to their $8010$ counterparts at twice the modulation frequency and compared also to a static RPP pattern. Many orders of magnitude reduction during 200 ps of simulation time are observed. We expect the advantages in the proper parameter regimes ($L_{HS} \approx L_{INT} \approx L_{spike}/2; \nu_{IAW} / ( 2 c_s k_0) \approx 0.1 $) to be less than three orders of magnitude which is largely sufficient to move the field from an LPI-plagued state at present, such as on NIF, to an LPI controlled state by properly designed, adapted and implemented STUD pulses. In addition, temporal  interlacing of STUD pulses on crossing beams will allow the control of inter-beam energy transfer which is lacking at present in both direct and indirect drive settings. Crossed beam energy transfer can neither be turned on nor off at present, nor can it be selectively targeted to subsets of beams. Static wavelength shifts between cones of beams are the only external knobs being exercised to date. STUD pulses can markedly improve on this. Additionally, for IFE, STUD pulses make a very strong case for the adoption of Green laser light where the implementation of the STUD pulse program is much less technically challenging from the laser damage threshold and conversion efficiency points of view and  will allow up to twice the total energy to be available for driving implosions. Conversion to Green or even a Blue laser ICF or IFE program without adequate control of LPI could be seen ultimately as being foolhardy. 

This work was supported in part by a grant from the DOE NNSA SSAA program and a Phase I SBIR from DOE OFES and the DOE NNSA-OFES Joint HEDP Program. 
Part of the simulations have be performed on the facilities of IDRIS-CNRS, Orsay, France.
\begin{figure}
 \begin{minipage}{6.9cm}
\resizebox{.97\columnwidth}{!}{%
\centering\includegraphics{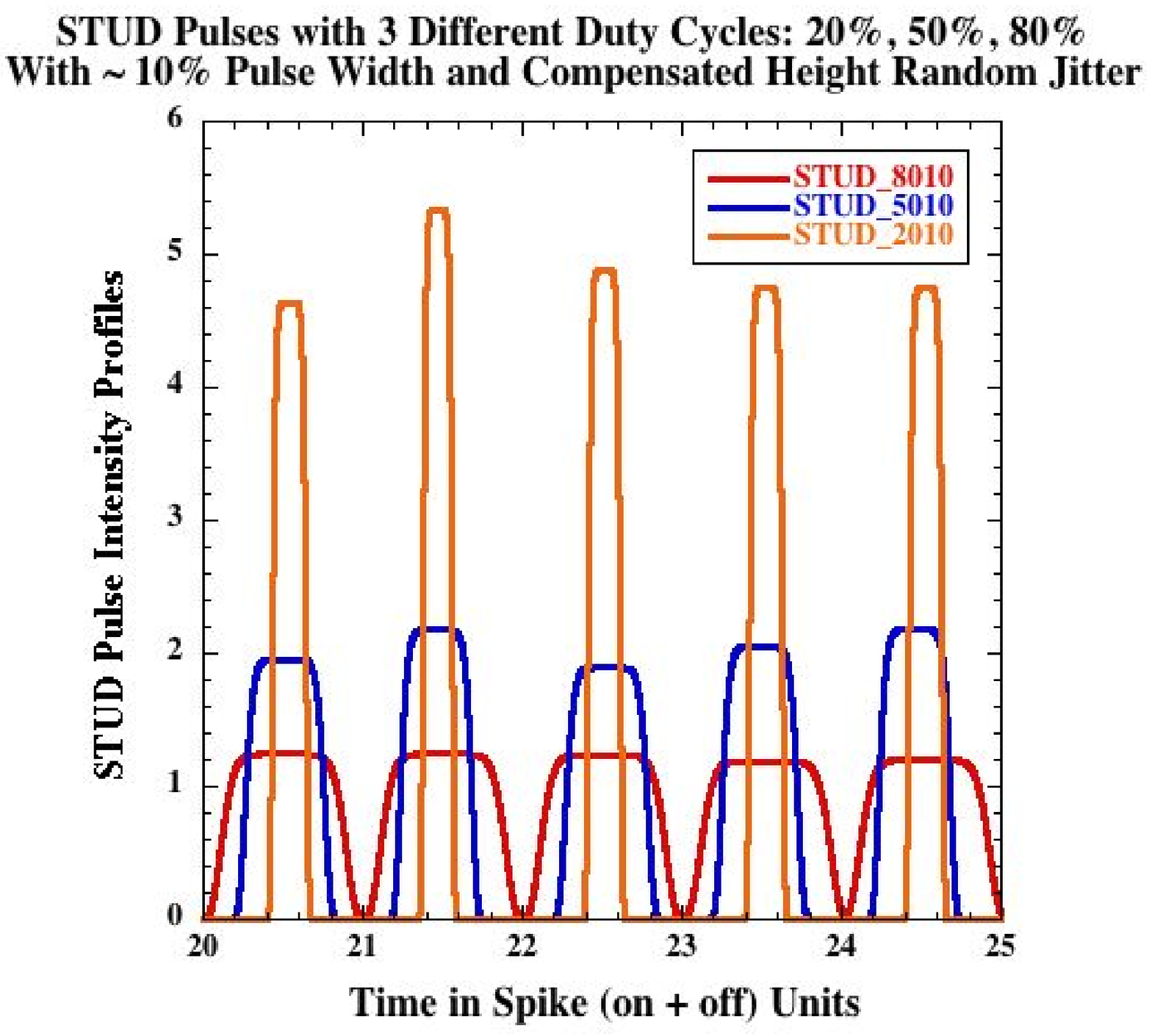}}
\caption{Samples of three different STUD pulses are shown with 20\%, 50\% and 80\% duty cycle and 10\% jitter in the widths and heights of the spikes such that their product is constant for each spike (on, off) pair.  }
\label{fig1}
 \end{minipage} \hfill
 \begin{minipage}{6.7cm}
\resizebox{.91\columnwidth}{!}{%
  \centering\includegraphics{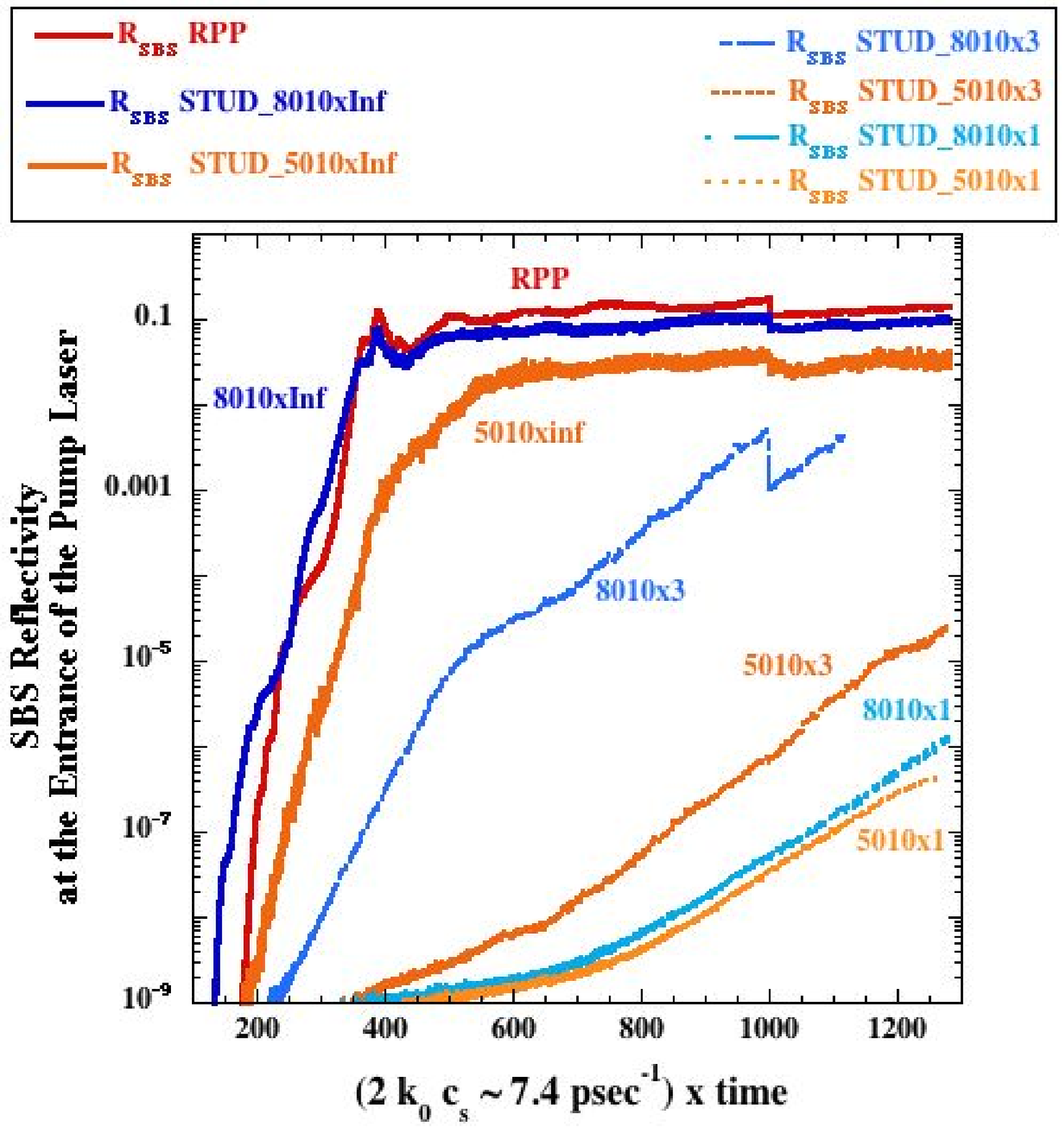}}
\caption{SBS Reflectivity as a function of time for
different cases:
STUD $\!5010\!\!\times\!1, \!\times3, \!\times\infty$ are compared to STUD $\!8010\!\!\times\!1, \!\times3, \!\times\infty,$ with double the modulation frequency, and to a static RPP patterned beam.}
\label{fig2}  
 \end{minipage}
\end{figure}
%
%
%

\end{document}